Microstructure and Structural Phase Transitions in Iron-Based Superconductors

Wang Zhen（王臻）, Cai Yao（蔡瑶）, Yang Huai-Xin（杨槐馨）, Tian Huan-Fang（田焕芳）, Wang Zhi-Wei(王秩伟)， Ma Chao（马超）, Chen Zhen（陈震）and Li Jian-Qi（李建奇）†

*Beijing National Laboratory for Condensed Matter Physics, Institute of Physics, Chinese Academy of Sciences, Beijing 100190, China*

Abstract:

Crystal structures and microstructural features, such as structural phase transitions, defect structures, chemical and structural inhomogeneities, are known to have profound effects on the physical properties of superconducting materials. Recently, a large amount of works on the structural properties of Fe-based high-$T_c$ superconductors have been published. According to the major systems under study, this review article will mainly focus on those typical microstructural features from samples that have been well characterized by physical measurements. (a) Certain common structural features are discussed. In particular, crystal structural features for different superconducting families, local structural distortions in the $Fe_2Pn_2$ (Pn=P, As, Sb) or $Fe_2Ch_2$ (Ch=S, Se, Te) blocks, and structural transformations in the 122 system. (b) In FeTe(Se) (11-family), the superconductivity, chemical and structural inhomogeneities are investigated and discussed in correlation with superconductivity. (c) In $K_{0.8}Fe_{1.6+x}Se_2$ system, we focus on typical compounds with emphasis on Fe-vacancy order and phase separations. The microstructural features in other superconducting materials are also briefly discussed.

**Corresponding author**
LJQ@aphy.iphy.ac.cn

# 1. Introduction

The discovery of superconductivity in F-doped LaOFeAs [1] with $T_c$ = 26K following an earlier report of superconductivity in $LaFePO_{1-x}F_x$ [2] has stimulated a renewed interest in investigations of high-Tc superconductors. It is considered that materials containing iron have always been considered hardly to undergo superconducting transition at high temperature, because of antagonistic properties between superconductivity and magnetism. Soon after this discovery, a number of superconducting families with $T_c$ ranging from 2K to 56K have been obtained,[3-4] such as LnFeAsO (Ln = rare earth mental, such as La, Sm) with ZrCuSiAs structure (so called 1111 type), moreover, Sm substitution for La in 1111 structure can raise the $T_c$ up to 55K.[4] In addition to 1111 system, other families of Fe-based superconductors are found to show a rich variety of structural and physical properties, including $A_EFe_2As_2$ ($A_E$=alkali-earth metals) with $ThCr_2Si_2$ structure (the 122 type),[5, 6] AFeAs (A=Li, Na) with the $Cu_2Sb$ structure (the 111 type),[7-10] and FeSe (the 11 type).[11] Recently, experimental investigations also demonstrate that intercalation of additional layers this kind of layered systems could obtain new materials with superconductivity.[12-14] For example, $Sr_2VFeAsO_3$ (21113 type)[13] and $Sr_3Sc_2Fe_2As_2O_5$ (32225 type) [14] contain intercalated perovskite-like blocks and exhibit superconductivity at around 37K and 45K, respectively. Superconductivity in $K_yFe_{1.6+x}Se_2$ was discovered in 2010,[15] then experimental measurements in related systems show that this type of materials, e.g. $Rb_xFe_{2-y}Se_2$, $Cs_xFe_{2-y}Se_2$, $(Tl, K)_xFe_ySe_2$ and $(Rb, K)_xFe_ySe_2$, often shows the superconducting $T_c$ ranging from 28K to 32K depending on chemical compositions.[16, 17] The basic structures of $A_{0.8}Fe_{1.6+x}Se_2$ (A= K, Rb, Cs, Tl/K, Tl/Rb) superconductors show certain similar properties with the 122 system which has a space group of I4/mmm. In addition, the presence of Fe-vacancy ordering within the FeSe layer was observed commonly in this layered system, so it was called 122* type structure in previous literatures. Comparing with other Fe-based superconductor, the 122* system exhibits some remarkable physical and structural properties. For instance, the superconducting volume in most samples is only about 20% as estimated from the data of magnetic shielding obtained under high applied fields. Moreover, the 122* materials often show complex microstructural features arising from Fe-vacancy orders and phase separations. In this article, we will focus on discussions of crystal structures and microstructural features observed in a number of Fe-based superconducting systems, such as structural transformations, defect structures, chemical and structural inhomogeneities, these structural phenomena are considered to have profound effects on the physical properties of superconducting materials.

## 2. Some fundamental results on structures of Fe-based superconductors

The Fe-based superconductors in general adopt layered structures, the essential layers, i.e. the $Fe_2Pn_2$ (Pn=P, As, Sb) or $Fe_2Ch_2$ (Ch=S, Se, Te) blocks, play a critical role for the appearance of superconductivity in this kind of materials. Experimental investigations demonstrate that these blocks could stack up in a number of different ways in superconducting materials. For instance, intercalated layers (A) between $Fe_2Pn_2$ layers following a sequence of ...[$Fe_2Ph_2$]/A/[$Fe_2Ph_2$]/A/[$Fe_2Ph_2$]... can results in 122-type superconductors as commonly discussed in previous investigations. Figure 1 shows a series of structural models, illustrating schematically basic crystal structures of observed superconducting families. The quasi-two-dimensional $Fe_2Ph_2/Ch_2$ block consists of a square lattice of iron atoms with Pn/Ch atoms coordinating at the apical sites. High-resolution transmission electron microscopy (TEM) observations along the relevant directions can clearly reveal the atomic structural features in all structural layers. Figure 2 exhibits a high-resolution TEM image taken from a $SrFe_2As_2$ crystal along the [100] zone-axis direction, illustrating the visible atomic layers stacking along the c-axis direction.[18]

The FePh/Ch layers in the Fe-based compounds, as discussed similarly for the $CuO_2$ plane in Cu-based superconductors, also play a key role for understanding the superconductivity. The geometric structural features of FePh/Ch blocks are essentially in correlation with the magnetic orders and electronic properties of these materials due to the strong bonding between Fe-Fe and Fe-Ph/Ch. The ground state of Cu-based superconductors has a long-range antiferromagnetic (AFM) order with the Cu spins aligned as shown in Fig.3(a). [19, 20] It is notable that certain Fe-based superconductors also exhibit a similar magnetic order in the a-b basic plane, as illustrated in Fig.3(b), in which the spin of neighboring Fe ions arranged ferromagnetically along one chain and antiferromagnetically along the other direction in the orthorhombic structure, this type of magnetic structure is called collinear AFM order. The magnetic structure along c-axis could be either ferromagnetic or antiferromagnetic order depending on the crystal structure and intercalated layers. [21, 22] On the other hand, theoretical calculations and experimental measurements demonstrated that the magnetic ground state of FeTe has a bi-collinear AFM order as presented in Fig. 3(c). [23]

It is also noted that superconductivity and magnetic orders in Fe-based compounds could evidently affected by local structural distortion which can be tuned through chemical substitution and element doping as well. Doping at different sites in 122 and 1111 systems could eventually suppress the structural distortion and related spin density wave (SDW) order, [24] and then lead appearance of superconductivity. For instance, a small fraction of Co substitution for Fe could result in rich physical and structural properties in $BaFe_{2-x}Co_2As_2$. Figure 4(a) shows a phase diagram clearly illustrating the transition of AFM state into superconductivity following with Co-doping on the Fe sites.[25] For $FeTe_{1-x}Se_x$ compounds (11 system), FeTe has a tetragonal structure and shows up evident monoclinic distortion accompanying with a SDW state at low temperatures as shown in Fig. 4(b). [26] With the Se substitution for Te, the crystal structure at low temperatures changes from monoclinic to orthorhombic lattice which could be in favor of superconductivity. It is commonly noted that the maximum superconducting critical temperature is reached only when the structural distortion and magnetic order are completely suppressed.

Actually, the effects of local structural features on the superconductivity in Fe-based superconductors have been extensively investigated in recent investigations, and certain theoretical discussions and experimental results have been summarized in previous literatures.[27-29] It is widely accepted that the maximum $T_c$ favors an undistorted Pn(Ch)FePn(Ch) octahedron with ideal Fe-Pn(Ch) bond angle of 109.47°. Figure 5(a) shows the experimental results obtained in a variety of superconducting systems, in which the superconducting $T_c$ versus the Pn(Ch)-Fe-Pn(Ch) angle are plotted.[30] On the other hand, the superconducting $T_c$ can be raised with the increase of the height from Pn(Ch) to Fe plane ($h_{Pn/Ch}$) in the same block. Figure 5(b) shows the experimental results obtained from a few typical superconducting systems, demonstrating the evident effects of $h_{pn}$ on the superconductivity in Fe-based superconductors.[28-30] Moreover, Ogino's group reported that $T_c$ in iron pnictides can be increased slightly with the thickness rise of Fe-Pn(Ch) blocks but saturates at around a thickness of ~ 1.5nm.[30]

## 3. Structural phase transition in 122-system

The $A_EFe_2As_2$ ($A_E$= Sr, Ca, Ba) materials in general show visible structural transitions at low temperatures, anomalies of electrical resistivity and magnetic susceptibility in 122 materials are identified to be associated with the SDW instability and tetragonal (I4/mmm) to orthorhombic (Fmmm) (T-O) phase transition at low temperatures.[6, 21, 31] We herein illustrate the structural changes in correlation with the T-O phase transition as directly observed in $BaFe_2As_2$ and $CaFe_2As_2$ compounds by means of in-situ cooling TEM experiments from RT down to 20K.[18]

The $SrFe_2As_2$ compound undergoes a structural transition at about 250K, associated with a SDW instability.[21] As can be seen in Fig.6(a-c), bright-field image and corresponding selected area electron diffraction (DPs) show the typical structural changes through the T-O phase transition, illustrating the twinning lamellae in low-temperature orthorhombic phase which is heavily twinned with the (110)$_{orth}$ twinning plane. Different space groups are used in present studies to describe the two structures: **a** and **b** axis directions in the low-temperature orthorhombic phase refer to the [110] and [-110] directions in comparison with the room-temperature tetragonal phase. The width of the twin domains ranges from 100nm to 400nm at 100K, and the twin density could affected by thermal process, impurity content and sample preparation technique as we discussed in Ref. [32]. In Fig. 6(d), we propose a schematic structural model of the mirrored twinning structure in the Fe layer. The spots splitting along $<110>_{orth}$ direction with an angle ~0.1° in the DPs obtained from several twin domains in Fig.6(c) indicates the ratio of the lattice parameter a/b to be about 1.009, which is fundamentally in agreement with the data of the X-ray and neutron diffractions 1.010.[33] The relationship between the magnetic state and the twinning structure widely existing in the low temperature orthorhombic phase is an open issue under investigations.

Similar to $SrFe_2As_2$ samples, the $CaFe_2As_2$ compounds also undergo the T-O phase transition at about 170 K and show clear twin domains in the orthorhombic phase, while a new pseudo-periodic structural modulation with a wavelength about 40nm is observed in $CaFe_2As_2$ compounds at room temperature. As shown in Fig.6(e), the modulation waves commonly coexist in many crystals along two equivalent [110] and [1-10] directions of the tetragonal phase, and the two orthogonal modulations overlap to yield an interesting pattern called 'tweed' structure as observed in alloys [34] and Fe-doped $YBa_2Cu_3O_{7-x}$ superconductor.[35] In the low-temperature orthorhombic phase as shown in Fig. 6(f), the coexistence of twin lamella and tweed structure in general yields complex microstructure features in the a-b plane of $CaFe_2As_2$. It should be noted that tweed structures show no visible changes even below T-O transition, so we conclude that the twinning structures occur within Fe layers while tweed structures in the Ca layers. The dissimilarities in both electrical resistivity and magnetic susceptibility between the $SrFe_2As_2$ and $CaFe_2As_2$ are likely to be induced from this pseudo-periodic structural modulation.

The correlation between the SDW and superconductivity, as shown in the phase diagram of Fig. 4(a) for a typical 122 system, is considered being a fundamental issue concerned in Fe-based superconducting materials. In contrast with the 11 system, element doping in superconducting compounds of 1111 and 122 systems often results in structural inhomogeneities and defect structures. TEM observations on Ca$_{0.5}$Sr$_{0.5}$Fe$_{1.74}$Co$_{0.26}$As$_2$ superconductor [36] suggest that inhomogeneous distribution of Sr and Ca could be responsible for the relatively wider superconducting transition often observed in this kind of superconducting materials. Recently, measurements of spatially resolved spectroscopy on NaFe$_{1-x}$Co$_x$As$_2$ near the SDW and superconducting phase boundary directly reveals the presence of both SDW and superconducting gaps at the same atomic location, providing compelling evidences for the microscopic coexistence of the SDW and superconducting states.[37]

## 4. Defect structures in FeTe$_{1-x}$Se$_x$

The FeTe compound, as a typical material in the 11-system, exhibits a structural transition from tetragonal to triclinic phase at low temperature, and a clear AFM transition appear at around 65K.[38] Superconductivity can been introduced by Se substation for Te as discussed in above context (see the phase diagram of Fig. 4(b)), The Fe$_{1+y}$Se superconductor (T$_c$ ≈ 8K) also shows up visible structural transitions with lowering temperature, crystal lattice of this compound has a space group of P4/mmm at room temperature and undergoes a structural transition to an orthorhombic lattice with space group of Cmma due to structural distortions of FeSe$_4$ tetrahedra below 90K.[39] Moreover, it is noted that FeSe compounds has a complex phase diagram as discussed in Refs.[40,41], as a result, impurity phases, such as α-FeSe, Fe$_7$Se$_8$, Fe$_3$O$_4$ and Fe, often appear in the Fe$_{1+y}$Se samples.[42]

Measurements of physical properties reveal that all FeTe$_{1-x}$Se$_x$ samples are tetragonal paramagnetic metals at high temperatures, on the other hand, this kind of materials show a rich variety of physical phenomena at low temperatures, i.e. the SDW ordering for $x < 0.1$, spin glass-like static magnetic ordering for $0.1 < x < 0.3$ and superconductivity for $x > 0.3$.[26] One of the notable phenomena in this system is the presence of a spin glass-like state between the SDW and superconducting region, which is considered to be accompanied by the local structural distortion and lattice disorders.[43] Structural measurements using scanning transmission electron microscopy (STEM) and electron energy loss spectroscopy (EELS) on Fe$_{1+y}$Te$_{1-x}$Se$_x$ crystals demonstrated the presence of phase separation and chemical inhomogeneity on nano-scales.[44] TEM investigation on FeTe$_{0.7}$Se$_{0.3}$ samples show that the areas with fluctuation of Te concentration around 20% commonly exist in the superconducting compounds. Figure 7 show a series of experimental data obtained from a well-characterized superconducting sample. Fig. 7(b) shows the EELS data taken from line-scanning measurements as indicated in the Fig.7(a), and the fluctuation of Te concentrations matches well with the simultaneous high angle annual dark field (HAADF) image intensity profile in Fig.7(c), reflecting the Z-contrast alternations following with chemical fluctuation. The alterations of the Fe L$_3$/L$_2$ intensity ratio in EELS spectra also show a clear correlation with the fluctuation of Te concentration, as illustrated in fig.7(d), the Te-rich (poor) regions corresponds to lower (higher) L$_3$/L$_2$ intensity ratios. Careful analysis on these experimental data suggests the alternations of the d-electron occupancy on the Fe ions in different areas, implying a local variation in the magnetic moment on nano-scales. Moreover, the EELS measurements demonstrate that the valence states of Te and Se could deviate from the nominal valence state of +2, and the substitution of Se with Te leads to a charge transfer and/or an effective doping as discussed in Ref. [45].

In addition, other kinds of defect structures, such as twinning domains and grain boundaries, which could affect the superconductivity and critical current density ($J_c$), have also been examined in Fe$_{1+y}$Se films by using scanning electron microscopy (STM).[46] Twin boundaries aligned with 45° to the Fe-Fe bond directions were found to noticeably suppress the superconducting gap as shown in Fig. 8. The variation in superconductivity is likely caused by the local structural changes in the areas neighboring to twin boundaries. It is also noted in structural investigations of the iron-based superconductors that the superconductivity $T_c$ depends remarkably on the height from Se to Fe layer ($h_{Se}$), and reaches a maximum value at $h_{Se} \approx$ 1.38Å, and decreases abruptly away from this value. In FeSe, $h_{Se}$ ranges from 1.38 Å to 1.45 Å nearby the twin boundary, and the increase of $h_{Se}$ could suppress $T_c$ at twin boundaries. Moreover, experimental measurements show that magnetic vortices are preferentially pinned around twin boundaries due to the visible changes of local structure and degradation of superconductivity. [47]

## 5. Phase separation and structural transition in $K_{0.8}Fe_{1.6+x}Se_2$

Intercalation of alkali metals into the 122 layered system not only yields superconducting in $A_yFe_{1.6+x}Se_2$ but also results in Fe vacancies in the FeSe layer. Our recent TEM observations demonstrate that certain microstructure phenomena and local structural inhomogeneities are essentially in correlation with the appearance of Fe-vacancies in this layer, such as structural modulations from different configuration of vacancy orders, order-disorder transitions and phase separations in superconducting materials. These microstructural features could fundamentally affect the physical properties, especially the magnetic orders and superconductivity.[48-50]

### 5.1 Structural modulations and Fe-vacancy ordering in $K_yFe_{1.6+x}Se_2$

The Fe vacancy order and correlated AFM state in this system have been firstly studied in 1987 by means of neutron diffraction on a material with nominal composition of $TlFe_2S_2$, two kinds of Fe-vacancy orders were found but no superconductivity was observed[51]. Based on structural analysis in $(Tl, K)Fe_xSe_2$ materials, the Fe-vacancy orders are respectively discussed for samples with x= 1.5, 1.6, and 2.0.[17] In recent investigations of $A_yFe_{1.6+x}Se_2$ superconductors, microstructural features and superstructure modulations have been extensively studied by means of TEM, STEM,[48, 50, 52] XRD[53] and neutron scattering,[49, 54-55] the typical superstructures in correlation with Fe-vacancy orders have been addressed, e.g. the $\sqrt{5}\times\sqrt{5}$ superstructure ($q_1$-modulation), $\sqrt{2}\times\sqrt{2}$ superstructure ($q_2$-modulation) and $\sqrt{2}\times2\sqrt{2}$ superstructure ($q_3$-modulation) have been identified as different structural phases in $A_yFe_{1.6+x}Se_2$ system.

Based on the structural data obtained from well-characterized samples, we first discussed fundamental structural features for the $q_1$- and $q_3$- modulations. As shown in Fig. 9, our TEM observations and diffraction measurements on $K_yFe_{1.6+x}Se_2$ reveal that the wave vectors for these two modulations can be written as $q_1=1/5(3, 1, 0)$ and $q_3=1/4(3, 1, 0)$ in the notion of tetragonal 122 structure.[48, 57] The chemical composition of $q_1$-phase is estimated to be $K_{0.8}Fe_{1.6}Se_2$ (so called 245 phase) with one Fe vacancy of every five Fe atoms, this phase has a space group of I4/m with the lattice parameters $a_1=b_1=\sqrt{5}a_s$, and $c_1=c_s$, where sublattice parameters $a_s= 3.913$ Å and $c_s = 14.10$ Å. Similarly, the chemical composition of $q_3$-phase can be written as $KFe_{1.5}Se_2$ (so called 234 phase), Figure 10(b) shows a structure model for $q_3$-phase which has a space group of Ibam with lattice parameters of $a_3=2\sqrt{2}a_s$ and $b_3=\sqrt{2}a_s$. Figure 9 exhibits electron diffraction patterns and high-resolution TEM/STEM images taken from thin crystals for $q_1$- and $q_3$-modulated phases. In Fig. 9(d) and (e), the Fe-vacancy orders with visible periodic features can be clearly read out. Moreover, the $q_1$-modulation and related Fe-vacancy order in $TlFe_{1.6}Se_2$ has been directly observed by using an aberration-corrected STEM as shown in Figure 9(f),[57] this STEM image was taken from a thin area along the [310] zone-axis direction in which the Fe-vacancy can be clearly recognized as dark dots separated by four iron columns. It is also noted that Fe ions in both $K_2Fe_4Se_5$ and $K_2Fe_3Se_4$ phases has a valence state of $Fe^{2+}$, therefore these two phases are expect to exhibit semiconducting/insulating properties in agreement with the experimental results.

The magnetic structures in association with different Fe-vacancy orders, including $q_1$- and $q_3$-phase, have been discussed by theoretical calculations.[58] Neutron diffraction [49] on $K_2Fe_4Se_5$ crystal shows an antiferromagnetic transition at $T_N \approx 559$ K below the Fe-vacancy order-disorder transition at $T_S \approx 578$ K. The magnetic unit cell as depicted in Fig. 10(a) is the same as the $K_2Fe_4Se_5$ crystal unit cell with a low occupancy on one Fe site surrounded by a square of four fully occupied Fe sites. The Fe magnetic moments form a collinear AFM structure with the c-axis being the easy magnetic axis, the magnetic moment of Fe (~3.31 $\mu_B$) obtained at 11K is the largest value among all iron pnictide and chalcogenide superconductors. Other neutron diffraction studies on $Cs_yFe_{2-x}Se_2$ and $Rb_yFe_{1.6+x}Se_2$ also revealed the similar phase transition and AFM structure.[54] Magnetic ordering associated with the Fe-vacancy order have been also observed in the $KFe_{1.5}Se_2$ phase[55], as clearly illustrated in Fig. 10(b). The magnetic moment of Fe in present case is estimated to be about 2.8 $\mu_B$ compatible with the theoretical data.[58]

## 5.2 Phase separation in $A_{0.8}Fe_{1.6+x}Se_2$ superconducting system

The phase separation phenomenon in $A_{0.8}Fe_{1.6+x}Se_2$ superconductors have been directly observed by using TEM,[48, 50] scanning electron microscopy (SEM), [59, 60] STM,[61] X-ray diffraction[62, 63] and neutron scattering,[54] careful structural analysis demonstrate that notable microstructural features in present system are fundamentally arising from the phase separations in both micro- and nano-scales. Furthermore, Mössbauer spectroscopy,[64] Raman spectroscopy [65] and angle resolved photoemission spectroscopy[66] also give strong evidences for the presence of phase separation in this superconducting system, in which there are about 80-90% volume of the sample exhibiting antiferromagnetic order and a volume fraction of about 10-20% could attribute to superconductivity at low temperature.

Direct observations of the atomic structural features in the phase separated state have been firstly performed by using high-resolution TEM in $K_yFe_{1.6+x}Se_2$ superconductors.[48, 50] Remarkable microstructure phenomena have been clearly observed, such as the Fe-vacancy ordering, structural inhomogeneous and the phase-separation layers along the c-axis direction. As shown in Fig.11, the SADPs and HRTEMs for $K_{0.8}Fe_{1.6+x}Se_2$ superconductors reveal the coexistence of the $K_2Fe_4Se_5$ phase and another $q_2$ phase on nanometer scale, this phase separated phenomenon can been clearly observed in a-b plane and along c-axis. The superstructure spots in the SADPs show the $q_2$-phase has a modulated vector of $q_2 = 1/2(1, 1, 0)$. Further study suggests that this $q_2$ phase is in correlation with the superconductivity in present system.

In order to clearly view the microstructure features in phase separated states for this layered system, a series of well-characterized $K_{0.8}Fe_{1.6+x}Se_2$ single-crystal samples were prepared for structural investigations.[50] Measurements of resistivity and magnetic susceptibility show that $K_{0.8}Fe_{1.6}Se_2$ exhibits semiconducting property in low-temperature range. With the increase of Fe content, the $K_{0.8}Fe_{1.6+x}Se_2$ compounds exhibit superconductivity for x larger than 0.06. Transition between the AFM phase $K_{0.8}Fe_{1.6}Se_2$ and the superconductors is likely in association with a mixed-phase in which obvious structural inhomogeneity can be often observed. In our experimental investigations, it is noted that SEM observations under the backscattering mode unambiguously reveal visible alterations of chemical compositions on $K_{0.8}Fe_{1.76}Se_2$ superconducting crystals at the micron/nano-scales,[59] and rich- and poor-Fe regions can be evidently distinguished in the SEM secondary electron images. Careful analysis suggests that the most remarkable microstructure features in present system is the appearance of micro-scale stripe-like patterns in all superconducting samples as clearly shown in Fig. 12 and the density of the stripes increases progressively with the increase of Fe content. These stripe patterns also show visible correlation with the anisotropic assembly of superconducting domains along the [110] and [1-10] directions in the a-b plane. Moreover, the 3-dimeisonal structural features for phase separated states has been also examined, a structural framework arising from the stacking of the macro-stripes can be clearly recognized along the [113] directions. As presented in Fig.12, the Fe-poor domains are dominant by the known antiferromagnetic 245 phase with $q_1 = 1/5(3, 1, 0)$ and superstructure phase with a modulation of $q_2 = 1/2(1, 1, 0)$ is mostly concentrated in the Fe-rich stripe areas. The $q_2$-phase with composition of $K_{0.5}Fe_2Se_2$, with an expanded c-axis compared with 245 phase, is considered to be the superconducting phase,[50, 62] which is thought to be induced by the ordered arrangement of K vacancy. In addition to the notable stripe structures, the nano-scale phase separation mentioned in previous literatures also appears around and within the stripe area in the superconducting materials.

The phase separation nature in other related materials has been also studied. For instance, high-resolution electron backscatter diffraction (HR-EBSD) mapping on $Cs_xFe_{2-y}Se_2$ have been used to analysis the local variations in composition and lattice parameter on micron scale in the phase separated state.[60] The experimental results demonstrate that the phase separated state containing a Fe-rich minor phase distributes as a plate-like morphology throughout the crystal in the matrix of $q_1$-superstructing phase, which is inconsistent with our experimental results. Figure 13 show the HR-EBSD maps for $Cs_{0.8}Fe_{1.9}Se_2$ and $Cs_{0.87}Fe_{2.04}Se_2$ with $T_c$ = 27K and 23K respectively, illustrating relatively larger c/a ratio in the minor phase than that of the matrix. X-ray study on this sample found a 5% increase of c/a ratio corresponding to 2.4% expanded c-axis and 2.4% contracted a-axis.[67] The authors explained these changes as a result of further nano-scale phase separation within the Fe-rich minor region, this microstructure phenomenon was also observed in our TEM study.

Recently, structural studies on $K_{0.8}Fe_{1.6+x}Se_2$ superconductors using high-resolution synchrotron and single-crystal x-ray diffraction[63] identified four distinct phase: the well-known semiconducting $K_2Fe_4Se_5$ phase ($q_1$ phase), a metallic superconducting phase $K_xFe_2Se_2$ ($q_2$ phase) with x ranging from 0.38 to 0.58, an insulator phase $KFe_{1.6}Se_2$ with no vacancy ordering and an oxidized $K_{0.51}Fe_{1.7}Se_2$. Based on these results, they also state that additional Fe in $K_2Fe_4Se_5$ could induce a distinct $K_xFe_2Se_2$ precipitate, which is in consistent with the dots/stripe picture as observed in our microstructure observations. Due to the complexity of the $K_{0.8}Fe_{1.6+x}Se_2$ superconductors, certain structural and physical properties for either the parent phase or superconducting phase are still in debate. Measurements of scanning tunneling microscopy and spectroscopy on $K_{0.8}Fe_{1.6+x}Se_2$ superconductors suggested a stoichiometric mental $KFe_2Se_2$ with $\sqrt{2}\times\sqrt{2}$ charge ordering as the parent compound [68], and superconductivity is induced in $KFe_2Se_2$ by either Se vacancies (e.g. superconducting $KFe_2Se_{2-z}$ with Se vacancies ) or interacting with the AFM $K_2Fe_4Se_5$ compound (superconducting $KFe_2Se_2$ with $\sqrt{2}\times\sqrt{5}$ charge ordering). NMR measurements [69] conclude that the superconducting phase might have the formula $A_{0.3}Fe_2Se_2$ (A=K, Rb).

## 5.3 Phase transitions in $K_2Fe_4Se_5$ and $K_{0.8}Fe_{1.6+x}Se_2$ superconductors

Neutron diffraction measurements on a $K_2Fe_4Se_5$ single crystal demonstrated that the $K_2Fe_4Se_5$ phase exhibits a AFM order in correlation with an Fe-vacancy ordering transition at about 558K.[49] In order to directly reveal the alteration of microstructure in association with Fe-vacancy ordering, we performed a series of in-situ TEM and X-ray diffraction experiments on $K_2Fe_4Se_5$ samples [56] in the temperature range from 300K up to 800K. Figure 14(a) shows schematically the changes of resistivity and magnetic susceptibility for $K_yFe_{2-x}Se_2$, at high temperatures as extensively discussed in Ref. [70], the $K_2Fe_4Se_5$ phase undergoes an AFM transition following the Fe-vacancy ordering and often shows visible resistivity anomalies at high temperatures. Figure 14(b) exhibits a series of selected-area electron diffraction patterns taken along [001] zone-axis direction taken from a $K_2Fe_4Se_5$ crystal, illustrating the visible changes of the superstructure with the increase of temperature. This in-situ TEM observation clearly demonstrate that the $K_2Fe_4Se_5$ phase contains a notable superstructure modulation ($q_1$) at 300K and undergoes a transformation, via an intermediate phase with short-rang order, towards a Fe-vacancy disordering state. The diffused superstructure spots observed above 500K suggesting a short-range order possibly resulting from an intermediate state short coherent length as similarly discussed in SrNdCaCuOF.[71] The high-temperature structure can be well characterized by a 122 structure with space group I4/mmm with lattice parameters of a = b = 3.913Å and c = 14.10Å. Moreover, our powder x-ray diffraction on a $K_2Fe_4Se_5$ sample also revealed clear structural changes in the temperature range from 300K to 600K as shown in Fig. 14(c). It is recognizable that the peaks corresponding to superstructures marked by asterisks disappear at temperatures over 600K. These results are in good agreement with what observed in-situ TEM investigations.

Phase separation in $K_{0.8}Fe_{1.6+x}Se_2$ superconductors as an important issue has been also observed using in-situ synchrotron radiation in the transmission mode,[72] demonstrating a clear phase separation between two competing phases: the majority $q_1$ phase with an in-plane expanded lattice and a minority non-magnetic $q_2$ phase with a compressed in-plane lattice. The high-temperature structure above 600K is found to be tetragonal 122-type as discussed in above context. As shown in Fig.15, the $q_1$ phase appears at 580K, in consistence with our in-situ TEM and neutron diffraction results, with further decrease of temperature, a new $q_2$ phase which merges with the $q_1$ phase appears at 520K, The two phases are assigned to a lattice-electronic instability, $K_{0.8}Fe_{1.6}Se_2$ is considered as a typical system tuned at a Lifshitz critical point with an electronic topological transition that gives a multi-gaps superconductor.[72]

## 6. Summary

In this article, we have demonstrated the complex microstructural phenomena in the Fe-based high-$T_c$ superconductors. The richness and complexity of the microstructural phenomena are truly remarkable. We have focused on the structural phase transitions, defect structures, chemical and structural inhomgeneities that commonly appear in a variety of Fe-based superconducting families. In particular, crystal structural features for different superconducting systems, local structural distortion in the $Fe_2Pn_2$ (Pn=P, As, Sb) or $Fe_2Ch_2$ (Ch=S, Se, Te) blocks, and structural phase transition. In FeTe(Se) (11-family), the superconductivity, chemical and structural defects are investigated in the related materials. Local structural analysis on the twinning boundaries clearly reveals the effects of structural changes on superconductivity. In $A_{0.8}Fe_{1.6+x}Se_2$ (a=K, Rb) superconducting materials, phase separated states on the micrometer and nanometer scales have been evidently observed. The $q_2$ phase with a composition of $K_yFe_2Se_2$ (y≈0.5) is considered to be the superconducting phase.

## Acknowledgement

This work was supported by National Basic Research Program of China 973 Program (Grant Nos. 2011CBA00101, 2010CB923002, 2011CB921703, 2012CB821404), the Natural Science Foundation of China (Grant Nos. 11274368, 51272277, 11074292, 11004229, 11190022), and Chinese Academy of Sciences.

# References

[1] Kamihara Y, Hiramatsu H, Hirano M, Kawamura R, Yanagi H, Kamiya T and Hosono H 2006 *J. Am. Chem. Soc.* **128** 10012

[2] Kamihara Y, Watanabe T, Hirano M and Hosono H 2008 *J. Am. Chem. Soc.* **130** 3296

[3] Chen X H, Wu T, Wu G, Liu R H, Chen H and Fang D F 2008 *Nature* **453** 761; Takahashi H, Igawa K, Arii K, Kamihara Y, Hirano M and Hosono H 2008 *Nature* **453** 376; Kito H, Eisaki H and Iyo A 2008 *J. Phys. Soc. Japan* **77** 063707; Wang C, Li L J, Chi S, Zhu Z W, Ren Z, Li Y K, Wang Y T, Lin X, Luo Y K, Jiang S, Xu X F, Cao G H and Xu Z A 2008 *Europhys. Lett.* **83** 67006

[4] Ren Z A, Lu W, Yang J, Yi W, Shen X L, Li Z C, Che G C, Dong X L, Sun L L, Zhou F and Zhao Z X 2008 *Chin. Phys. Lett.* **25** 2215

[5] Sefat A S, Jin R, McGuire M A, Sales B C, Singh D J and Mandrus D 2008 *Phys. Rev. Lett.* **101** 117004;   Rotter M, Tegel M and Johrendt D 2008 *Phys. Rev. Lett.* **101** 107006; Ronning F, Klimczuk T, Bauer E D, Volz H, Thompson J D, 2008 *J. Phys.: Condens. Matter.* **20** 322201

[6] Chen G F, Li Z, Li G, Hu W Z, Dong J, Zhang X D, Zheng P, Wang N L and Luo J L 2008 *Chin. Phys. Lett.* **25** 3403

[7] Wang X C, Liu Q Q, Lv Y X, Gao W B, Yang L X, Yu R C, Li F Y and Jin C Q 2008 *Solid State Commun.* **148** 538

[8] Pitcher M J, Parker D R, Adamson P, Herkelrath S J C, Boothroyd A T and Clarke S J 2008 *Chem. Commun.* **45** 5918

[9] Deng Z, Wang X C, Liu Q Q, Zhang S J, Lv Y X, Zhu J L, Yu R C and Jin C Q 2009 *Europhys. Lett.* **87** 37004

[10] Tapp J H, Tang Z, Lv B, Sasmal K, Lorenz B, Chu P C W and Guloy A M 2008 *Phys. Rev. B* **78** 060505(R)

[11] Hsu F C, Luo J Y, Yeh K W, Chen T K, Huang T W, Wu P M, Lee Y C, Huang Y L, Chu Y Y, Yan D C and Wu M K 2008 *Proc. Natl Acad. Sci.* **105** 14262; Sales B C, Sefat A S, McGuire M A, Jin R Y, Mandrus D and Mozharivskyj Y 2009 *Phys. Rev. B* **79** 094521

[12] Ogino H, Matsumura Y, Katsura Y, Ushiyama K, Horii S, Kishio K and Shimoyama J I 2009 *Supercond. Sci. Technol.* **22** 075008; Ogino H, Shimizu Y, Ushiyama K, Kawaguchi N. Kishio K,
18

Figure captions:

**Fig. 1.** Crystal structures for Fe-based superconductors with the 11, 111, 122, 1111, 21113 and 32225 type of structures. The FeCh/FePh (Pn: pnictogen, Ch: chalcogen) tetragonal blocks are highlighted in these structural models.

**Fig. 2.** High-resolution TEM image for SrFe$_2$As$_2$ taken along the [100] zone-axis direction, illustrating the layered structural features, a theoretical simulated image is superimposed for comparison. [18]

**Fig. 3.** Magnetic orders within the basic a-b plane for some parent phases: (a) La$_2$CuO$_4$, (b) BaFe$_2$As$_2$ and (c) FeTe. The chemical unit cells are marked as light green. The dark and light brown As/Te/Se atoms indicate their vertical positions above and below the Fe-layer, respectively. [20]

**Fig. 4.** Phase diagrams for iron-based superconductors of (a) Ba(Fe$_{2-x}$Co$_x$)$_2$As$_2$ and (b) FeTe$_{1-x}$Se$_x$. [25, 26]

**Fig. 5.** (a) Superconducting T$_c$ vs. the Pn(Ch)-Fe-Pn(Ch) angle in a number of systems. (b) Superconducting T$_c$ vs anion height (h$_{Pn/Ch}$) in iron pnictides and iron chalcogenides blocks. [30]

**Fig. 6.** Structural changes accompanying with the T-O transition for 122 parent compounds. Bight-field images for SrFe$_2$As$_2$ sample at (a) 300K and (b) 100 K with a twinning structure. (c) Electron-diffraction patterns taken along the [001] zone-axis direction at 100 K. Inset shows the enlarged (400) Bragg spot splitting along the <110>$_{orth}$ direction. (d) Structural model illustrating schematically twin boundaries within a Fe layer. TB represents twin boundary. (e) Bright-field TEM image of CaFe$_2$As$_2$ at room temperature, shows a tweed structure induced by pseudoperiodic modulation along <110>$_{tetra}$ direction. (f) Bright-field image for CaFe$_2$As$_2$ at 100 K

shows a mixture of the tweed and twin domains. [18]

**Fig. 7.** Phase separation and chemical inhomogeneity in a FeTe$_{0.7}$Se$_{0.3}$ sample. (b) The red spectrum shows the background subtracted EELS for the Te-M$_{4,5}$ edge along the line in the HAADF image of Fig.7(a), and the black spectrum shows a typical energy loss spectrum. (c) The black dotted curve shows the variation of Te content along the scanning line, and the histogram shows the HAADF intensity recorded simultaneously with the EELS data. (d) The black curve shows changes of the intensity ratio of Fe L$_3$/L$_2$ white-line. The red curve demonstrates that the Te content deviates from the average value. [44]

**Fig. 8.** STM data showing the suppression of superconductivity nearby a twin boundary in a FeSe film. (a) STM topography with a TB indicated by a white dashed line. (b) Atomically resolved topography of the TB. (c) Normalized dI=dV spectra taken at along the white solid line in (a). (d, e) Differential conductance maps recorded simultaneously with image (a) at energies of (d) zero and (e) 2.2 meV, respectively. [46]

**Fig. 9.** Electron diffraction patterns and high-resolution TEM images taken along relevant directions. (a) Electron diffraction pattern showing the q$_3$ = 1/4(3, 1, 0) modulation for K$_2$Fe$_3$Se$_4$, (b) Electron diffraction pattern showing the q$_1$=1/5(3, 1, 0) modulation for K$_2$Fe$_4$Se$_5$. (c) Electron diffraction pattern showing q1-modulation observed along [130] direction. HRTEM image illustrating the ordered arrangement of iron vacancies for 234 phase (d) and for 245 phase (e). (f) High angle annular dark field (HADDF) image taken along [310] zone-axis direction directly illustrates the iron-vacancy as the dark dots in consistence with the diffraction pattern in (e). [57, 48]

**Fig. 10.** Crystal and magnetic structures for (a) K$_{0.8}$Fe$_{1.6}$Se$_2$ and (b) KFe$_{1.5}$Se$_2$, Fe-vacancy orders and layered structural features are clearly illustrated. [49, 55]

**Fig. 11.** Phase separation between $q_1$-phase ($K_2Fe_4Se_5$) and $q_2$-phase with $q_2=1/2(1, 1, 0)$ on nano-scale in a $K_{0.8}Fe_{1.7}Se_2$ superconductor. The SADPs and corresponding HRTEM images taken from (a-b) [001] and (c-d) [310] zone-axis directions, illustrating the phase separation on nano-scale between $K_2Fe_4Se_5$ phase and $q_2$ phase. [48, 50]

**Fig. 12.** TEM/STEM images showing phase separation on micro-scale, the stripe pattern of the $q_2$ phase embedded in matrix of semiconducting $K_2Fe_4Se_5$ phase. SEM images of a-b plane in $K_{0.8}Fe_{1.6+x}Se_2$ samples with (a) x = 0, (b) 0.06, demonstrating the stripe emerges with the appearance of superconductivity as the Fe content increases. (c) STEM image of $K_{0.8}Fe_{1.7}Se_2$ along the c-axis direction shows the stripe pattern with darker contrast going along the [110] and [110] directions. SAEDPs taken from a light A area (d) with modulated vector of q1 = 1/5[a* +3b*] and the stripe B area (e) with modulation of $q_2$ = 1/2[a* + b*]. Elemental mapping of the rectangular area for (f) K element and (g) Fe element, respectively, illustrating the higher Fe concentration for the $q_2$ phase. [50]

**Fig. 13.** HR-EBSD analysis shows variation in c/a ratio for samples with nominal composition (a) $Cs_{0.8}Fe_{1.9}Se_2$ (Tc~27K) and $Cs_{0.87}Fe_{2.04}Se_2$ (Tc~23K). (c-d) Regions with c/a>3.9 refer to the minority phase. [60]

**Fig. 14.** (a) Schematic illustration of alterations in resistivity and magnetic susceptibility for $K_{0.8}Fe_{1.6}Se_2$ at high temperatures. The structural changes were observed in TEM and XRD measurements. (b) Electron diffraction patterns at different temperatures from 300K to 600K. (c) The XRD of $K_{0.8}Fe_{1.6}Se_2$ polycrystal samples at different temperature, the two arrows marks the (001) and (011) peaks of FeSe splitting at around 600K. [56]

**Fig. 15.** X-ray diffraction data of $K_{0.8}Fe_{1.6+x}Se_2$ shows the intrinsic phase separation.

(a) Temperature evolution of (200) peaks, the splitting of the peak at low temperature evident for the appearance of majority in-plane expanded $K_2Fe_4Se_5$ phase and the minority in-plane compressed $q_2$ superconducting phase. (b) Temperature evolution of the $q_1$ and $q_2$ superstructure satellite intensity during the heating and cooling runs. (c) The variation of the probability of the $q_2$ phase around the phase separation temperature and the temperature variation of the ratio between the intensity of $q_1$ satellites and basic diffraction spots. [72]

Fig.1

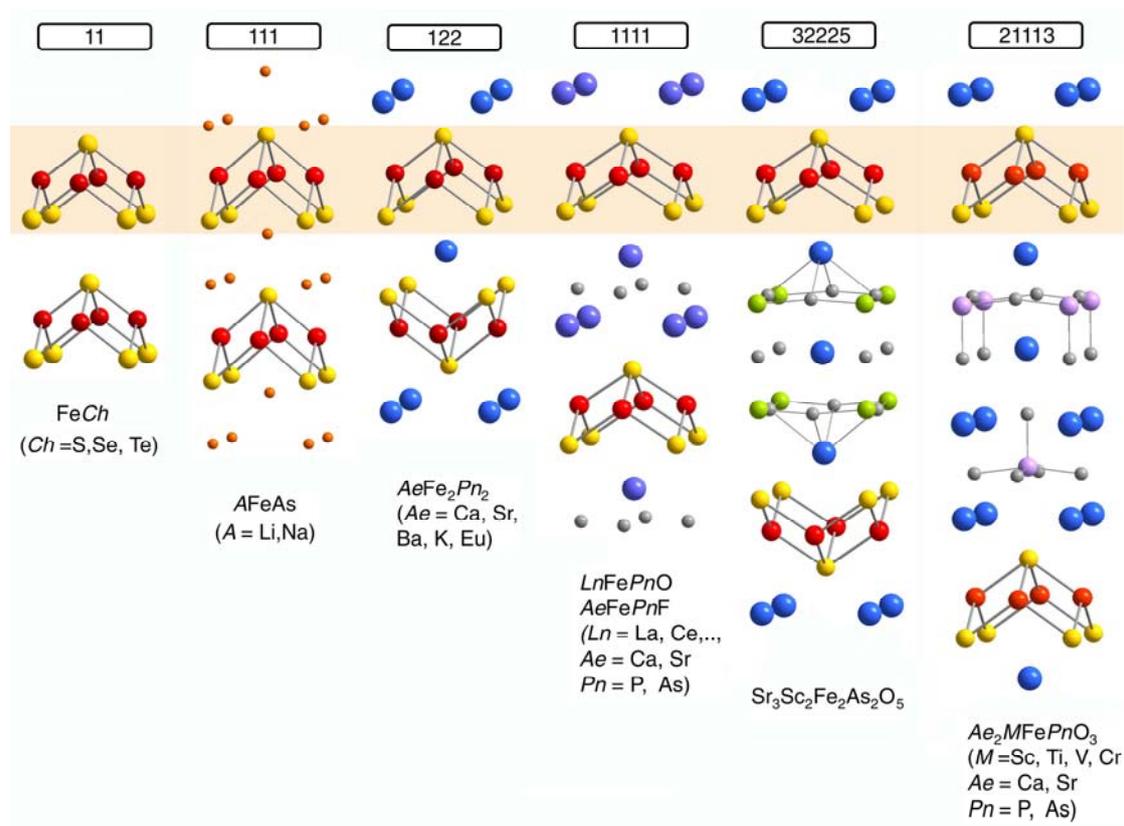

Fig.2

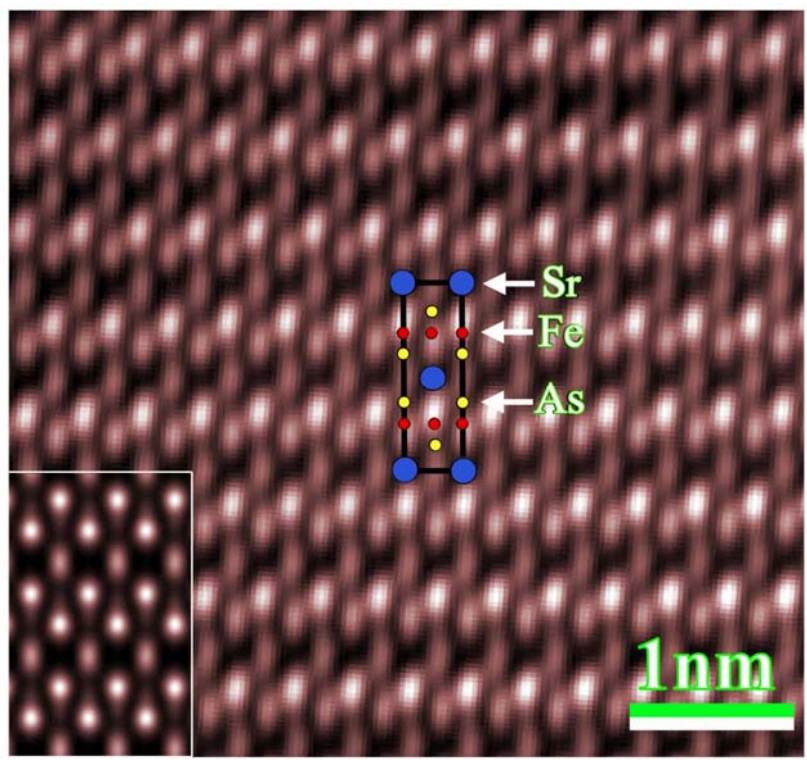

Fig.3

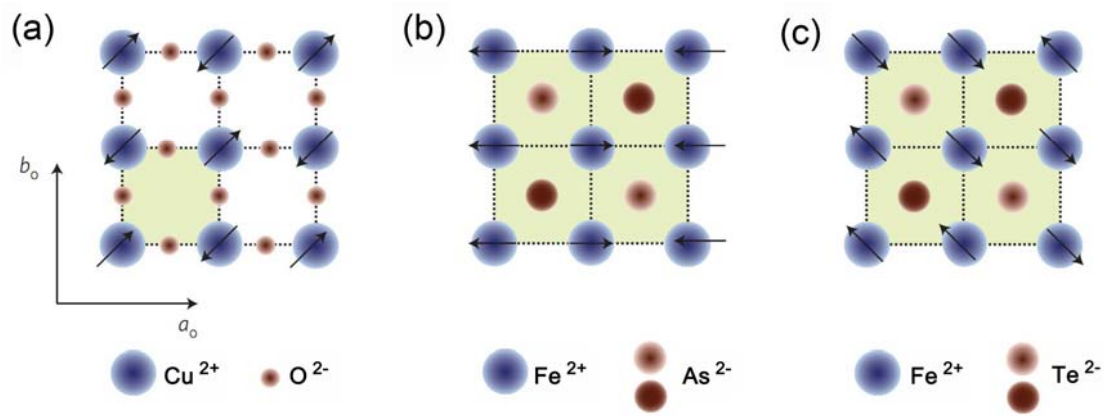

Fig. 4

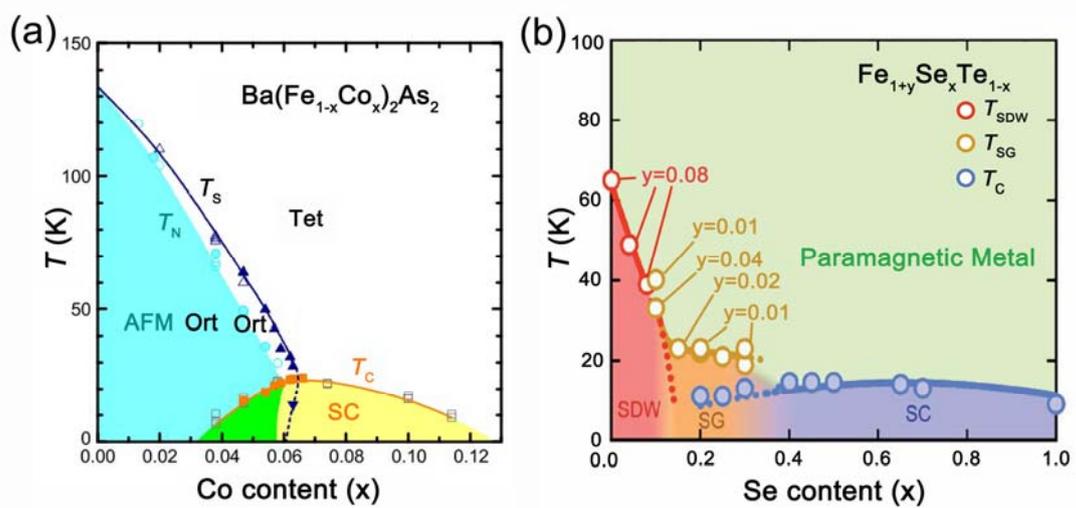

Fig. 5

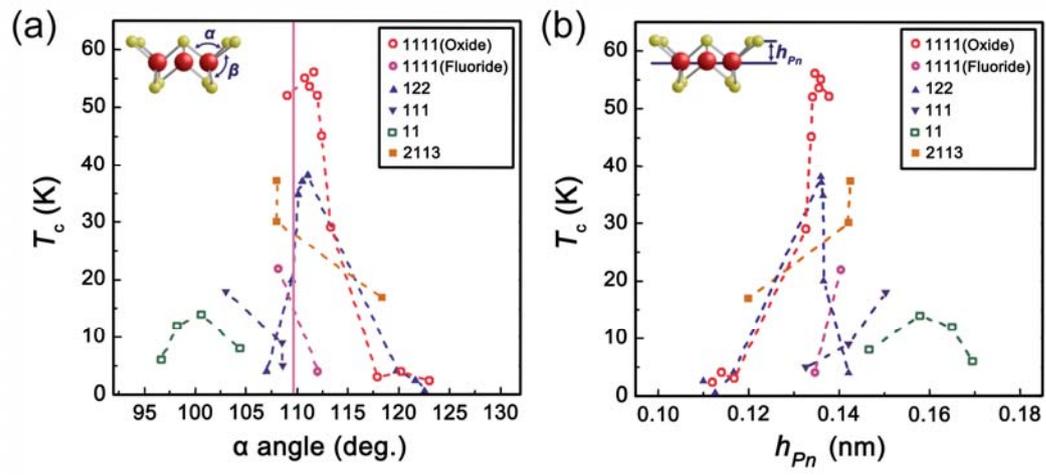

Fig. 6

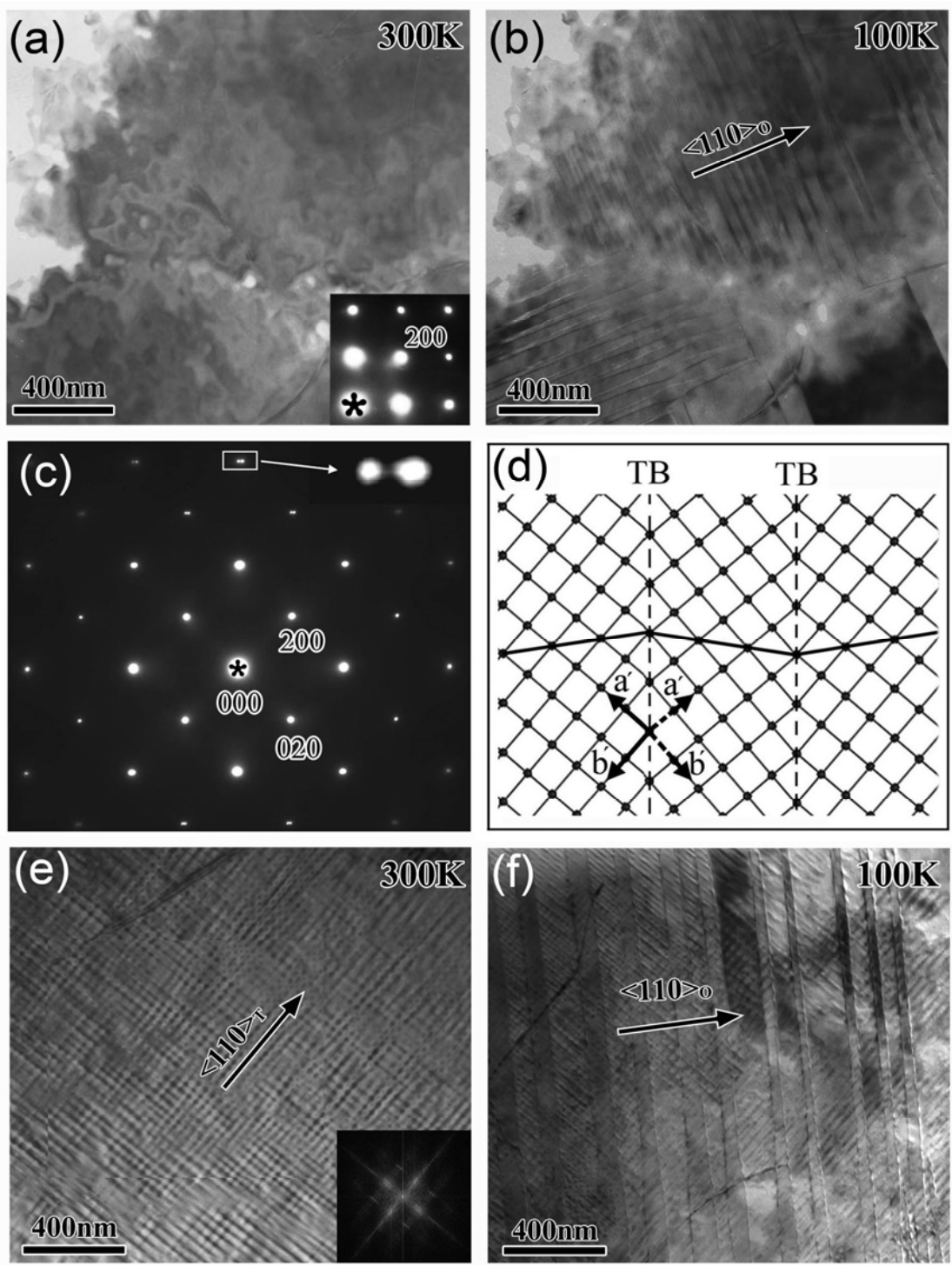

Fig. 7

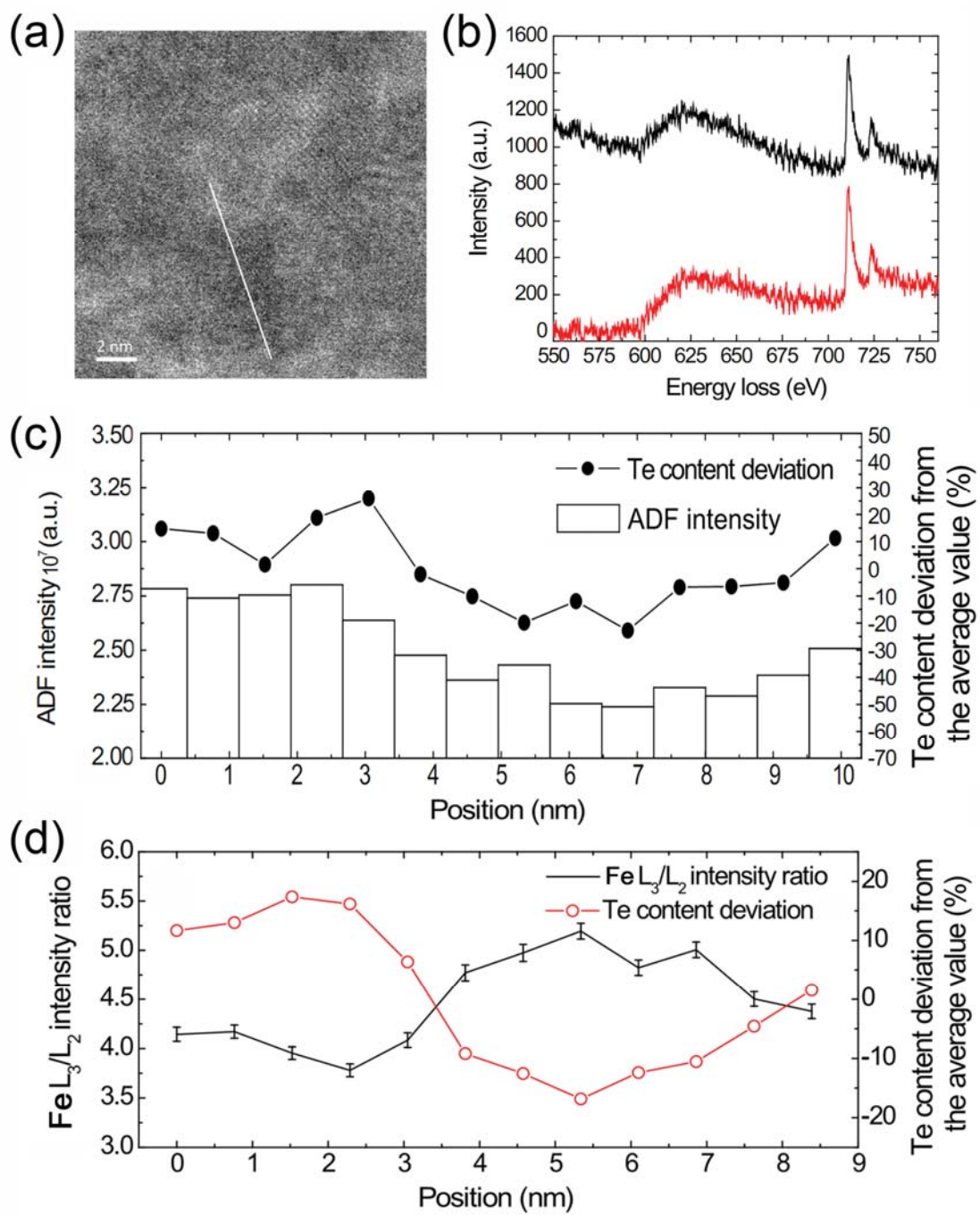

Fig. 8

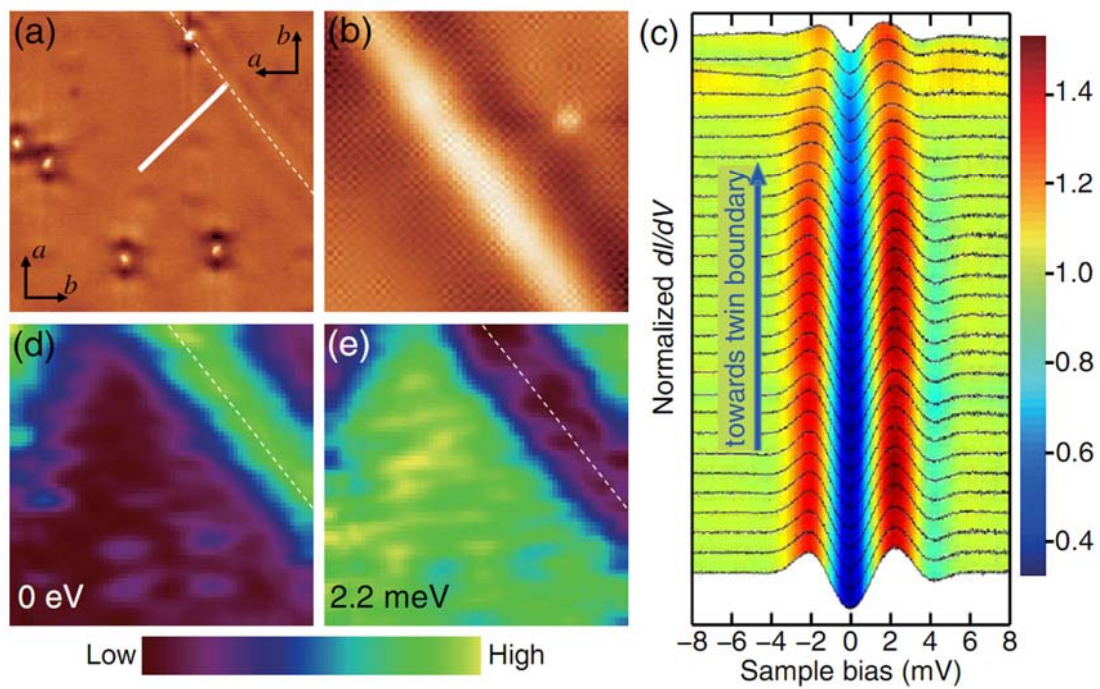

Fig. 9

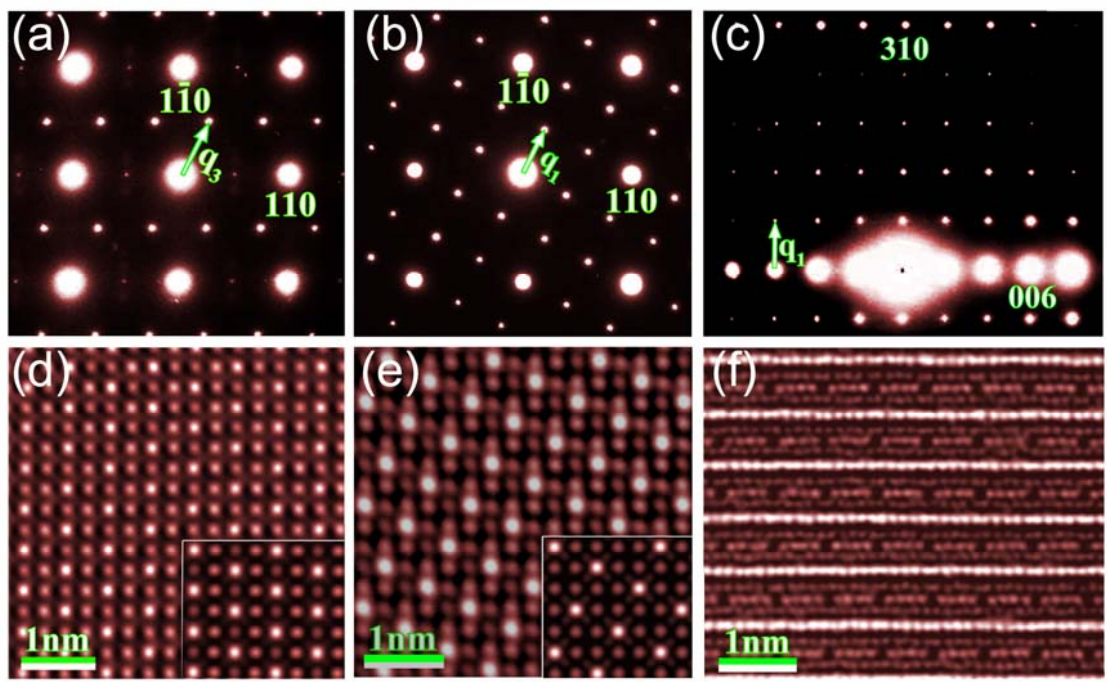

Fig. 10

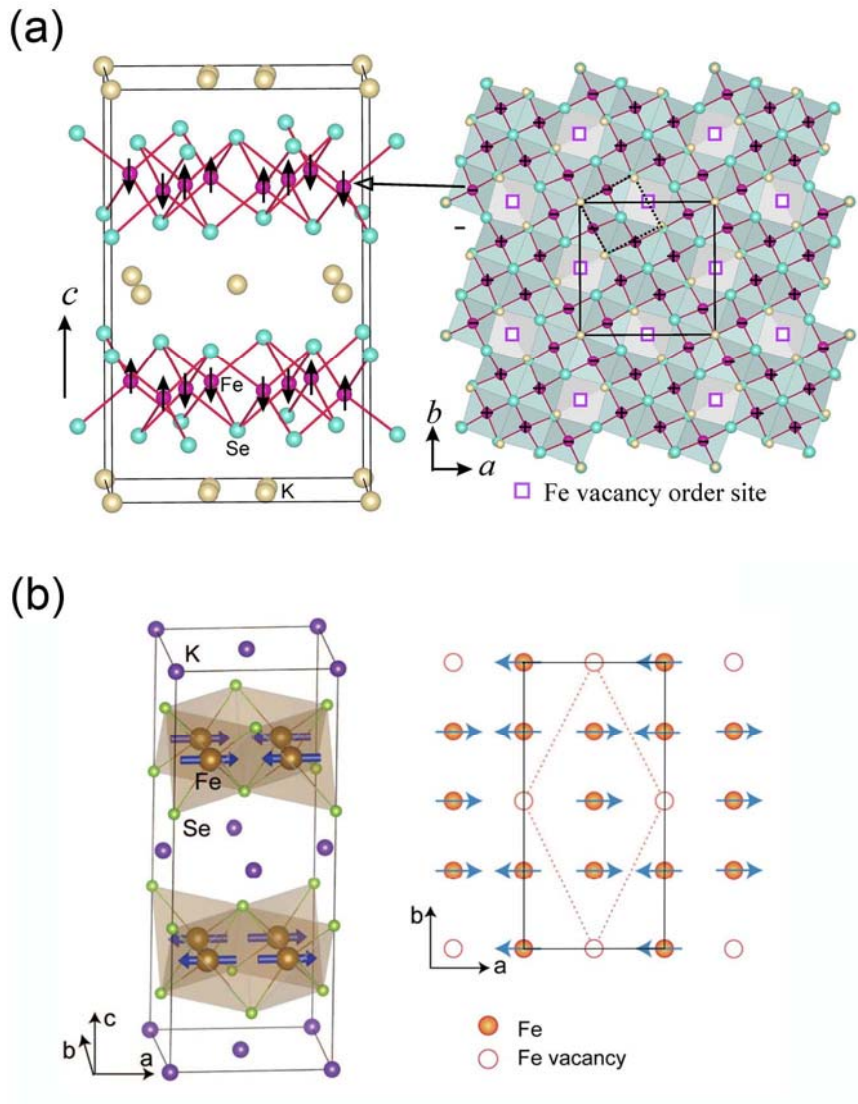

Fig. 11

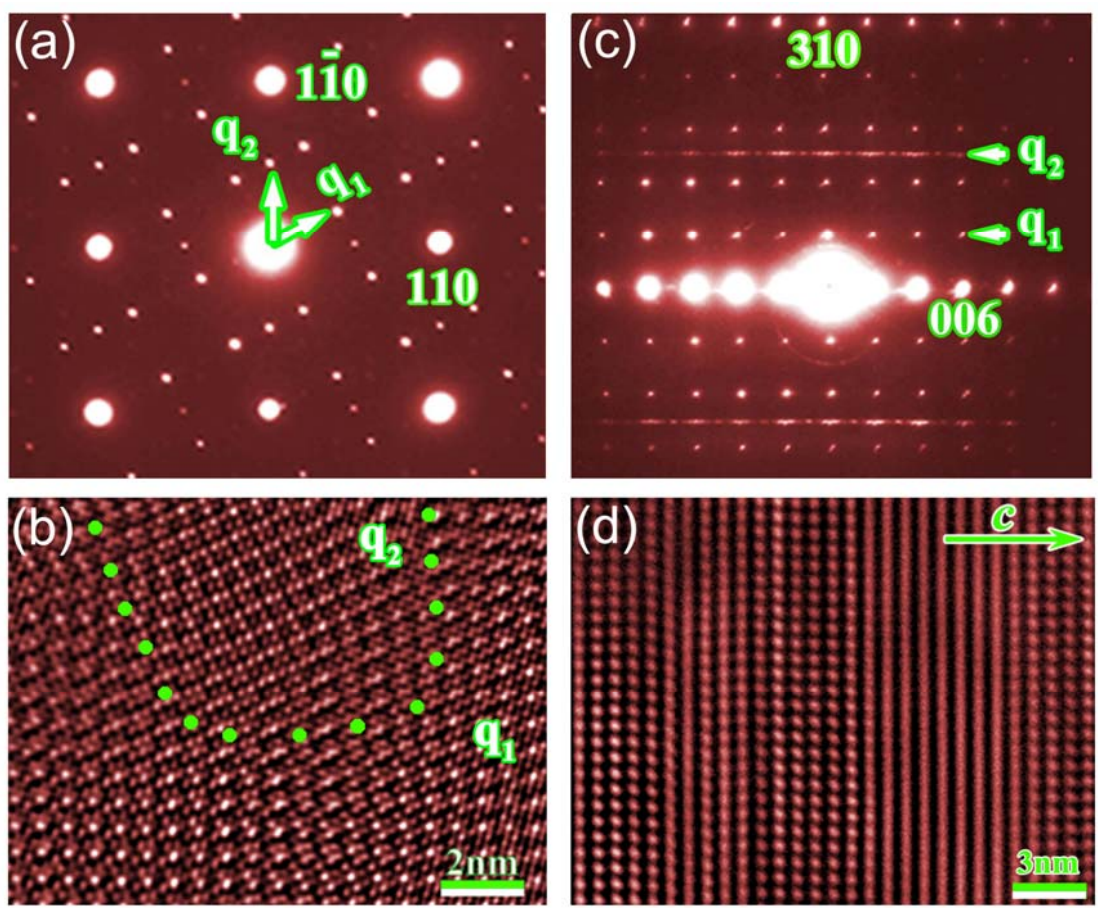

Fig. 12

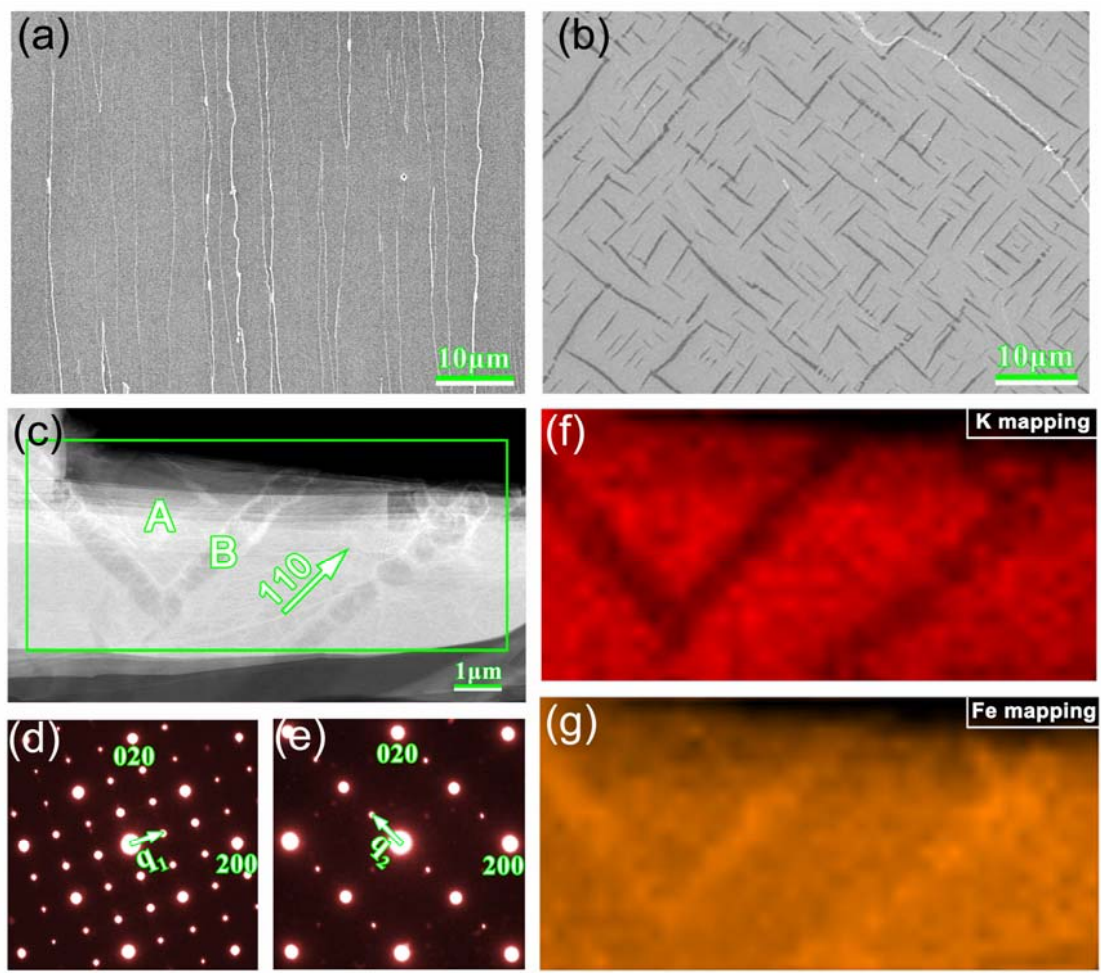

Fig. 13

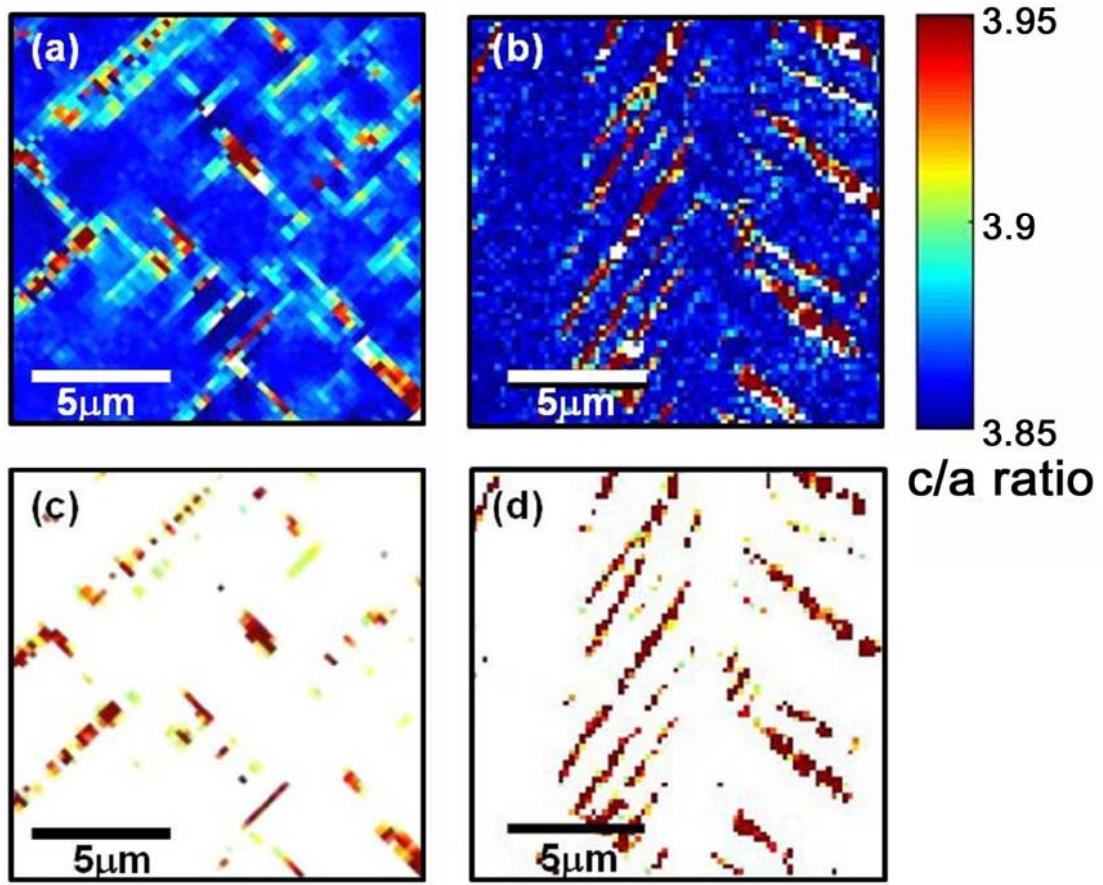

Fig. 14

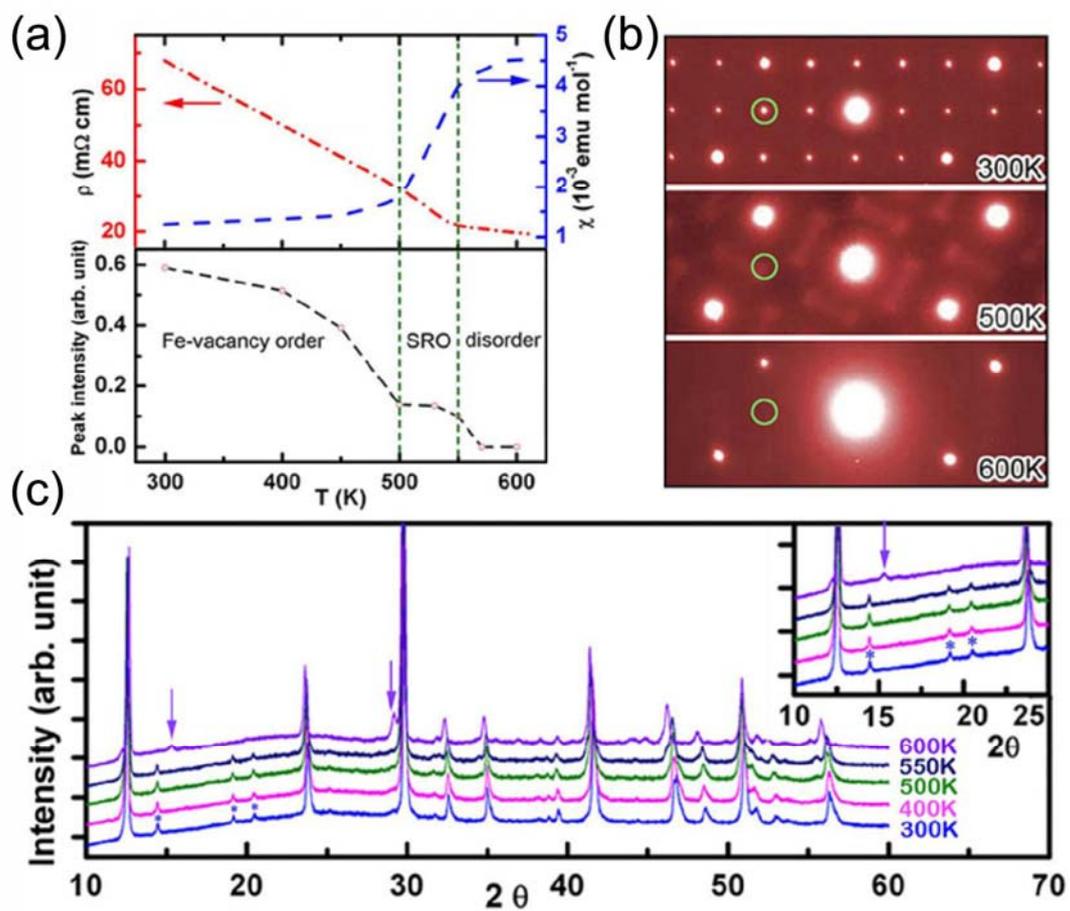

Fig. 15

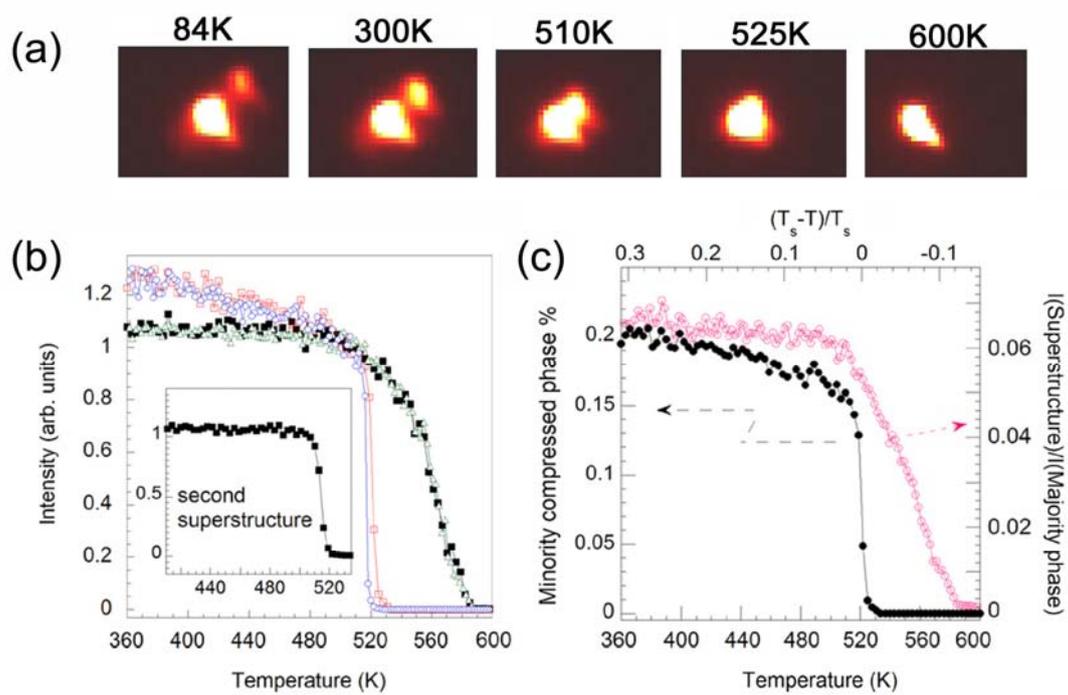